\documentclass[twoside]{article}
\usepackage{fleqn,espcrc2}

\usepackage{amssymb}

\newcommand{\cO}{{\cal O}}

\title{Current correlators and form factors in the resonance region
\thanks{Talk given at the 14th International Quantum Chromodynamics Conference, 7--12th July (2008), Montpellier (France). IFIC/08-54 report.}
}

\author{I.~Rosell \address{Departamento de Ciencias F\'{\i}sicas, Matem\' aticas y de la Computaci\' on, Universidad CEU Cardenal Herrera, c/Sant Bartomeu 55, E-46115 Alfara del Patriarca, Val\` encia, Spain}
\address{IFIC, Universitat de Val\` encia - CSIC, Apt. Correus 22085, E-46071 Val\` encia, Spain}}

\begin{document}

\begin{abstract}
\vspace{1pc}
Within Resonance Chiral Theory and in the context of QCD current correlators at next-to-leading order in $1/N_C$, we have analyzed the two-body form factors which include resonances as a final state . The short-distance constraints have been studied. One of the main motivations is the estimation of the chiral low-energy constants at subleading order, that is, keeping full control of the renormalization scale dependence. As an application we show the resonance estimation of some coupling, $L_{10}^r(\mu_0)=(-4.4\pm 0.9)\cdot 10^{-3}$ and $C_{87}^r(\mu_0)=(3.1\pm 1.1)\cdot 10^{-5}$.
\end{abstract}

\maketitle

\section{Motivation}

A possibility to deal with Quantum Chromodynamics (QCD) at low energies is the use of effective field theories \cite{EFT}. Chiral Perturbation Theory (ChPT) is the effective field theory of QCD at very low energies \cite{ChPT} and it is constructed by using a perturbative expansion in the momenta and masses of the pseudo-Goldstone bosons. Chiral Perturbation Theory has been used up to next-to-next-to-leading order and one of its major problems is the estimation of the increasing number of low-energy constants (LECs), once the desired precision is growing.

The construction of an effective lagrangian in the resonance region is much more involved. The existence of many resonances at intermediate energies ($M_\rho < E < 2$~GeV), the absence of a mass gap to integrate out the heavier degrees of freedom and the lack of a natural expansion parameter makes difficult a formal effective field theory approach. Resonance Chiral Theory (RChT) provides a correct framework to incorporate the resonance fields \cite{RChT}. One considers the most general possible lagrangian, including all terms consistent with assumed symmetry principles. The key ingredients that allows the phenomenology are the $1/N_C$ expansion and the use of the information coming from QCD.

Assuming confinement, the $N_C \rightarrow \infty$ limit guarantees that meson dynamics are described by tree-level interactions of an effective local lagrangian including only meson degrees of freedom, higher corrections in $1/N_C$ being obtained by loop corrections \cite{NC}. On the other hand, the use of the short-distance information from QCD is fundamental to reduce the number of unknown couplings.

It is important to stress that the only model dependence of our approach is the cut of the tower of resonances. In other words, although in the context of the $1/N_C$ expansion an infinite number of resonances is required to recover the usual QCD results, RChT only considers the lightest resonance multiplets. It seems a good approximation taking into account that heavier contributions are expected to be suppressed by their masses. Furthermore, this idea is supported by the phenomenology.

Besides the logic interest of resonance lagrangians to make physics in the resonance region, one of its main motivations is the estimation of the chiral LECs. That is, one can integrate out the resonance fields to estimate the chiral couplings in terms of resonance parameters. This matching is more interesting when the short-distance information is used to reduce the number of unknown parameters. Put differently, our phenomenological approach can be understood as a bridge between high and low energies: chiral LECs cannot be obtained directly from the high-energy lagrangian (ChPT is a non perturbative approach), but RChT is able to match with both theories. Somehow Resonance Chiral Theory allows to transport the information from high energy to ChPT.

This estimation has been made at leading order (LO) in $1/N_C$ for all the $\cO(p^4)$ and some $\cO(p^6)$ LECs \cite{RChT}. Actually, one of the main motivations to work with RChT at one-loop level is precisely this estimation at next-to-leading order (NLO), since the leading estimation is unable to control the renormalization-scale dependence of the couplings, which are unknown and could be sizable. This project is a step towards the understanding of quantum loops within RChT \cite{NLO,L10}, which is necessary also to improve the hadronic contributions to distinguish new physics effects from Standard Model results in some observables.

\section{Correlators and form factors}

Let us consider the two-point correlation function of two currents in the chiral limit:
\begin{eqnarray}
\Pi_{X}^{\mu\nu}\! &\equiv&  i \!\int \mathrm{d}^4x \, \mathrm{e}^{iqx}\;
\langle 0|T\left(J_X^\mu(x)J_X^\nu(0)^\dagger\right)|0\rangle \nonumber \\
&=& \left( -g^{\mu\nu} q^2 + q^\mu q^\nu \right)\,\Pi_{X}(q^2)\, , \nonumber \\
\Pi_{Y}\!&\equiv&  i\!\int \mathrm{d}^4x \, \mathrm{e}^{iqx}\;
\langle 0|T\left(J_Y(x)J_Y(0)^\dagger\right)|0\rangle \, , \label{def1}
\end{eqnarray}
where $J_X^\mu(x)$ can denote the vector or axial-vector currents and $J_Y(x)$ the scalar or pseudo-scalar densities.

At large $q^2$, the vector and axial-vector spectral functions tend to a constant whereas the scalar and pseudoscalar ones grow like $q^2$. Therefore, in the first case and considering that the spectral function is a sum of positive contributions corresponding to the different intermediate states and there is an infinite number of possible states, the absorptive contribution of a given state should vanish at infinite momentum transfer. Following the same argument, in the scalar and pseudoscalar case one would require spectral functions growing as a constant. However, the $SS-PP$ sum-rules, the Brodsky-Lepage counting rules \cite{brodsky-lepage} and the $1/q^2$ behavior of each one-particle intermediate cut seem to indicate that the vanishing assumption is reasonable.

Thus, we consider two sources of short-distance constraints. First, and taking into account that the optical theorem relates the spectral cuts with the corresponding two-body form factors, one can consider vanishing form factors. On the other hand, the matching between the resonance results with the ones obtained in the Operator Product Expansion (OPE) gives more constraints.

Within RChT we have calculated all two-body form factors associated with the scalar, pseudoscalar, vector and axial-vector currents \cite{L10} and have analyzed the high-energy constraints coming from well-behaved spectral functions.

\subsection{The V-A correlator in RChT}

We consider the difference between the two-point correlation function of two vector and two axial-vector currents, $\Pi(t)=\Pi_V(t)-\Pi_A(t)$, with $t=q^2$. Within RChT and at leading order, $\Pi(t)$ reads
\begin{equation}
\label{eq:LO}
 \Pi(t) =  \frac{2F^2}{t} +\sum_i \left[
 \frac{ 2\, F_{V_i}^2}{ M_{V_i}^2-t }
  - \frac{2\, F_{A_i}^2}{ M_{A_i}^2-t} \right] ,
\end{equation}
which involves an infinite number of vector and axial-vector resonance exchanges. At the NLO in $1/N_C$, $\Pi(t)$ has moreover one-loop corrections,
\begin{equation}\label{eq:Pi_structure}
\Pi(t) = \!
\frac{2 F^2}{t}  +\!  \sum_i \!\left[ \frac{2\, F_{V_i}^{r\,\, 2}}{M_{V_i}^{r\,\,2} - t}
-  \frac{2\, F_{A_i}^{r\,\,2}}{M_{A_i}^{r\,\, 2}  - t} \right] \!+  \widetilde{\Pi}(t) ,
\end{equation}
being $\widetilde{\Pi}(t)$ the contributions associated with two-meson absorptive cuts. We have considered only the lowest-mass two-particle exchanges: two pseudo-Goldstone bosons or one pseudo-Goldstone boson and one resonance (higher thresholds are kinematically suppressed \cite{L10}). $\Pi(t)$ can be obtained from the spectral functions through a dispersive relation, up to a term which at NLO has the same structure as tree-level resonance exchanges. Therefore, this term can be absorbed by a redefinition of the resonance couplings and masses.

We have adopted the Single Resonance Approximation (SRA) as a first approach, where just the lightest resonances with non-exotic quantum numbers are considered. Making use of the Weinberg sum-rules, the $S-P$ sum rules and well-behaved form factors, one is able to have all the relevant resonance parameters in terms of the pion decay constant $F$ and resonance masses \cite{L10}. The only remanent thing before going to the phenomenology is to fix $F_V^r$ and $F_A^r$, which can be done by studying the asymptotic behavior of the observable at hand at subleading order,
\begin{eqnarray}
F_V^{r\,\,2}&=&  \frac{F^2\,M_A^{r\,\,2}}{M_A^{r\,\,2}-M_V^{r\,\,2}}
       \! \left(1+\delta_{_{\rm NLO}}^{(1)}-\frac{M_V^2}{M_A^2}\delta_{_{\rm NLO}}^{(2)}  \right)
 ,\nonumber \\
F_A^{r\,\,2}&=&  \frac{F^2\,M_V^{r\,\,2}}{M_A^{r\,\,2}-M_V^{r\,\,2}}
       \! \left(1+\delta_{_{\rm NLO}}^{(1)}-\delta_{_{\rm NLO}}^{(2)} \right)
 ,\label{dmr}
\end{eqnarray}
where $\delta_{_{\rm NLO}}^{(i)}$ parameterize the asymptotic expressions of the one-loop contribution
\begin{eqnarray}
  \widetilde{\Pi}(t) \!\!\!
&=\!\!\!&  \frac{2F^2}{t}
\left( \delta_{_{\rm NLO}}^{(1)} +  \widetilde{\delta}_{_{\rm NLO}}^{(1)} \ln{\frac{-t}{M_V^2}}  \right)
 \nonumber \\
 &&+\frac{2F^2 M_V^2}{t^2} \!\left( \delta_{_{\rm
NLO}}^{(2)} \!+ \widetilde{\delta}_{_{\rm NLO}}^{(2)}
\ln\frac{-t}{M_V^2}\right)\!+\! \dots  ,
\end{eqnarray}
where the dots indicates subleading terms in the high-energy expansion. Note that the constraints $\widetilde{\delta}_{_{\rm NLO}}^{(1)}=\widetilde{\delta}_{_{\rm NLO}}^{(2)}=0$ give relations between masses, $M_A=M_V$ and $M_P=\sqrt{2}M_S$.

In order to avoid some incompatibilities between different short-distance constraints appearing in the Single Resonance Approximation \cite{L10}
, we have included additional resonance multiplets ($V'$ and $A'$). Obviously then there is much more couplings. Notwithstanding, we can use the known constraints coming from the $<VAP>$ Green-function analysis of Ref.~\cite{VAP} and group some new couplings into expected tiny correction $\epsilon_{i}$, given by
\begin{eqnarray}
\epsilon_{1}&=&\frac{F_{A'}^2}{F^2} -\frac{F_{V'}^2}{F^2} \,, \nonumber \\
\epsilon_{2}&=&\frac{F_{A'}^2 M_{A'}^2-F_{V'}^2 M_{V'}^2 }{F^2 M_V^2}\,, \nonumber \\
\epsilon_{3}&=&\frac{F_{V'} G_{V'}}{F^2}\, . \label{FVprime}
\end{eqnarray}
In our numerical calculations we only take the $\epsilon_i$ corrections into account when they appear at LO in $1/N_C$. All the procedure is equivalent to the case explained before, the only change is that now $F_V^r$ and $F_A^r$ also depends on $\epsilon_i$, so Eq.~(\ref{dmr}) reads now
\begin{eqnarray}
F_V^{r\,\,2}\!\!\!\!\!&=\!\!\!\!\!&\!  \frac{F^2\,M_A^{r\,\,2}}{M_A^{r\,\,2}\!-\!M_V^{r\,\,2}}
        \!\!\left[1\!+\!\epsilon_{1}\!+\!\delta_{_{\rm NLO}}^{(1)}\!
        \!-\!\frac{M_V^2}{M_A^2}\!\! \left(\!\epsilon_{2}\!+\!  \delta_{_{\rm NLO}}^{(2)}\!\right) \!\right]\!\!
 ,\nonumber \\
F_A^{r\,\,2}\!\!\!\!\!&=\!\!\!\!\!& \! \frac{F^2 \,M_V^{r\,\,2}}{M_A^{r\,\,2}\!-\!M_V^{r\,\,2}}
        \!\!\left[1\!+\!\epsilon_{1}\!+\!\delta_{_{\rm NLO}}^{(1)}\!\!-\!\epsilon_{2}\!-\!\delta_{_{\rm NLO}}^{(2)} \right]\!\!
 ,\label{dmrbis}
\end{eqnarray}
where of course the expressions of $\delta_{_{\rm NLO}}^{(i)}$ have changed \cite{L10}. The constraints $\widetilde{\delta}_{_{\rm NLO}}^{(1)}=\widetilde{\delta}_{_{\rm NLO}}^{(2)}=0$ are now used to fix $M_{V'}$ and $M_{A'}$.

All the complete expressions of present and next section are shown in Ref.\cite{L10}.

\section{The chiral couplings $L_{10}^r(\mu)$ and $C_{87}^r(\mu)$}

At very low energies $\Pi(t)$ is determined by ChPT:
\begin{eqnarray}
  \Pi(t) \!\!&\!\!=\!\!&\!\!  \frac{2 F^2}{t}  -  8 L_{10}^r(\mu)  - \frac{\Gamma_{10}}{4\pi^2} \left( \frac{5}{3}-\ln \frac{-t}{\mu^2} \right)\nonumber \\
&& \!\! + \frac{t}{F^2} \bigg[ 16C_{87}^r (\mu)  -\frac{\Gamma_{87}^{(L)}}{2\pi^2} \left( \frac{5}{3}-\ln \frac{-t}{\mu^2} \right)  \nonumber \\
&& \qquad  \qquad +\cO\left(N_C^{0}\right) \bigg]+  \cO\left(t^2\right)  ,\label{eq:Pi_chpt}
\end{eqnarray}
with $\Gamma_{10} = -1/4$ and  $\Gamma_{87}^{(L)} = - L_9/2$. The couplings $F^2$, $L_{10}$ and $C_{87}/F^2$ are of $\cO(N_C)$, while $\Gamma_{10}$ and $\Gamma_{87}^{(L)}/F^2$ are of $\cO(N_C^0)$ and represent a NLO effect.

\subsection{The large-$N_C$ limit in RChT}

The low-energy expansion of Eq.~(\ref{eq:LO}) determines, at LO in $1/N_C$, the chiral LECs appearing in Eq.~(\ref{eq:Pi_chpt}):
\begin{eqnarray}
L_{10}  &=&  - \frac{F_V^2}{4 M_V^2}  + \frac{F_A^2}{4 M_A^2}
 \approx -5.3 \cdot 10^{-3}  ,\nonumber \\
C_{87}&=&   \frac{F^2 F_V^2}{8 M_V^4}  - \frac{F^2 F_A^2}{8 M_A^4}
\approx 4.3 \cdot 10^{-5} \, ,
\label{eq:L10-LO}
\end{eqnarray}
where we have considered the SRA and have used the relations and inputs of Ref.~\cite{L10}.

\subsection{Next-to-leading order corrections}

The low-energy expansion of Eq.~(\ref{eq:Pi_structure}) determines now the couplings $L_{10}^r$ and $C_{87}^r$ at NLO in the $1/N_C$ expansion, with a control of the renormalization scale dependence. As it has been pointed out before, we have done the calculation under the SRA and including additional $V'$ and $A'$ \cite{L10}:
\begin{eqnarray}
 L_{10}^r(\mu_0)|^{\mathrm{SRA}} &=&  (-5.2 \pm 0.4 ) \, \cdot \, 10^{-3}\, , \nonumber  \\
 L_{10}^r(\mu_0)|^{\mathrm{V'A'}}&=& \left(-3.6 \pm 0.9 \right)\cdot 10^{-3} \, , \nonumber  \\
C_{87}^r(\mu_0)|^{\mathrm{SRA}}&=& (3.9\pm 0.6 )\, \cdot\, 10^{-5} \,, \nonumber \\
C_{87}^r(\mu_0)|^{\mathrm{V'A'}}&=& \left(2.2 \pm 1.1\right)\cdot 10^{-5}\,, \label{numbers}
\end{eqnarray}
being $\mu_0=770$~MeV.

\section{Conclusions}

Resonance Chiral Theory is an effective approach that allows to handle QCD at intermediate energies and is ruled by the $1/N_C$ expansion and constrained by the high-energy information. We have deeply investigated the constraints coming from well-behaved form factors in the spirit of correlators at NLO \cite{L10}.
\begin{table}
\begin{center}
\caption{Comparison with different estimations.} \label{comparison}
\begin{tabular}{|c|c|c|}
\hline
 & $10^3 \cdot L_{10}^r (\mu_0)$  & $ 10^5 \cdot  C_{87}^r (\mu_0)$    \\
\hline
 This work   &    $-4.4 \pm 0.9$ &
$ 3.1 \pm 1.1 $ \\
Ref.~\cite{bijnens-talavera}   &
$-5.5 \pm 0.7$
 &
 \\
Ref.~\cite{davier-girlanda}
&
$-5.13 \pm 0.19$
& \\
Ref.~\cite{martin} &
$ -4.10 \pm 0.29 $ &
$3.85 \pm 0.13$ \\
Ref.~\cite{peris} &
& $4.5 \pm 0.4$ \\
\hline
\end{tabular}
\end{center}
\vspace{-0.85cm}
\end{table}

One of the main motivations to deal with RChT at one-loop level is the estimation of the LECs, since tree-level predictions are unable to pin down the scale dependence of the ChPT couplings, which often are sizable. Here we have followed a general procedure described in Refs.~\cite{NLO,L10} to estimate the $\cO(p^4)$ and $\cO(p^6)$ couplings appearing in $\Pi(t)=\Pi_V(t)-\Pi_A(t)$, in terms of only the pion decay constant $F$ and resonance masses.

Combining the results of Eq.~(\ref{numbers}), obtained under the Single Resonance Approximation and including extra multiplets, one gets finally
\begin{eqnarray}
 L_{10}^r(\mu_0)&=& \left(-4.4 \pm 0.9\right)\cdot 10^{-3}  ,
\nonumber \\
C_{87}^r(\mu_0)&=& \left(3.1 \pm 1.1\right)\cdot 10^{-5} .
\end{eqnarray}
The general agreement with previous estimations is shown in Table~1. Note that our estimation is the only theoretical prediction at NLO.

Other chiral LECs can be estimated at next-to-leading order in $1/N_C$ by following the method explained here and applied to other observables. Work in this direction is in preparation.
\vspace{0.2cm}

\noindent {\bf Acknowledgments}  \\
I wish to thank S.~Narison for the organization of the conference and A.~Pich and J.J.~Sanz-Cillero for their helpful comments. This work has been supported in part by the Generalitat under grant GVPRE/2008/413, by the Spanish Government, under grants FPA2007-60323 and CSD2007-00042 (CPAN), and by the EU MRTN-CT-2006-035482 (FLAVIAnet).

\end{document}